# Topological Light Nodal Lines in FCC Lattice

Takuto Kawakami and Xiao Hu[*]

*International Center for Materials Nanoarchitectonics (WPI-MANA),
National Institute for Materials Science, Tsukuba 305-0044, Japan*
[*]e-mail: HU.Xiao@nims.go.jp

**Light cone with the speed of light independent of its wavelength in vacuum has been known for long time. In the present work, we unveil that in a face-centered-cubic (fcc) lattice of dielectric spheres novel light cones can be created over closed loops in momentum space, dubbed as nodal lines (NL), and that as a consequence of the nontrivial topology of NL interface states with a drumhead-shaped band structure appear where light can be slowed down significantly. We discuss that photonic pseudogaps found in previous experimental and theoretical studies for fcc photonic crystals are consistent with the present finding of NL. This offers a unique chance to confirm the existence of NL as a novel topological state.**

Recently, much attention has been paid to topological semimetals characterized by linear energy dispersions over closed loops in three-dimensional (3D) momentum space [1-13]. As a consequence of the nontrivial topology associated with a nodal line (NL), an interface state appears which exhibits a drumhead-shaped band [1]. The density of states (DOS) is enhanced significantly by this interface band, which is expected as an ideal platform for achieving high-temperature superconductors and other novel states with strong correlations [14]. Up to the moment of this writing, the novel topological state has not been confirmed experimentally yet.

On the other hand, light cone in vacuum has been well known for long time. The linear frequency dispersion relation, meaning that the light speed is constant irrespectively to its wave length, is a consequence of the absence of interaction. It has also become clear during the past few decades that when dielectric spheres are arranged in a regular pattern in 3D space, light can only propagate along directions permitted by the pattern symmetry with nonlinear dispersions, and in some cases light cannot propagate at all [15]. The term "photonic crystal" has been coined, where positive ions in a solid crystal are replaced by a regular arrangement of dielectric spheres with permittivity different from the environment [15,16].

In the present work, exploring the symmetry of the face-centered-cubic (fcc) structure we reveal that topological light cones can be created in a fcc lattice of dielectric spheres, where linear frequency dispersions are realized over loops in momentum space. We show explicitly that because of the nontrivial topology of light NL drumhead-shaped interface bands appear where light can be slowed down significantly [17-19]. With the similarity between fermionic and bosonic systems in generating nontrivial topology in mind, a concept proposed by Haldane and Raghu [20] and becoming the focus of subsequent researches (for a review see [21]), this provides a unique platform for confirming the existence of topological NL.

We notice that in the early days of research on photonic crystals efforts were devoted to quest for photonic band gap where light is stopped totally [22-24]. It turned out experimentally, however, that it is hard to achieve this goal with the fcc structure [22]. In theoretical works based on numerical calculations on Maxwell's equations, linear band crossings were obtained close to $U$ point and a two-fold degeneracy was found at *W* point preventing gap opening in the fcc structure [23, 24] (see also [16]). The DOS reduction at certain ***k*** points was known as photonic pseudogap. All these results are consistent with our finding of NL in fcc photonic crystals, which will be uncovered in what follows on the basis of analytic and numeric calculations with focus on topology supported by the unique symmetry of the fcc structure.

We consider a fcc lattice of dielectric spheres in background of another dielectric material as displayed in Fig. 1(a) with the first Brillouin zone (BZ) shown in Fig. 1(b). A dielectric sphere can accommodate electromagnetic modes similar to those of atomic *s*-, *p*-, *d*- and *f*-orbits. In fcc lattice, some of the orbits are preferred in accordance with the $O_h$ point group associated with the cubic structure: the three orbits $[p_x, p_y, p_z]$ correspond to the $T_{1u}$ representation and $[d_{xy}, d_{xz}, d_{yz}]$ to the $T_{2g}$ representation, which are our main concerns [25].

Of many symmetries inherent in the fcc structure, mirror symmetry is crucial for generating NL. To be specific, we consider the mirror operation with respect to the *xy* plane. The mirror symmetry of the present system is characterized by $M_{xy}H(k_x, k_y, -k_z)M_{xy}^{-1} = H(k_x, k_y, k_z)$, where $M_{xy}$ is the mirror reflection operator, $H(\boldsymbol{k})$ is the Hamiltonian describing orbital hopping among lattice sites as a function of momentum (see supplementary materials). On the $k_z = 0$ plane, where the commutation relation $[M_{xy}, H(\boldsymbol{k})] = 0$ is satisfied, the Hamiltonian is block diagonalized according to the eigenvalues of $M_{xy}$, $m_{xy} = \pm 1$, the parity upon the mirror reflection. It is easy to see that $p_x$ and $p_y$ orbits fall into the even-parity block whereas $p_z$ orbit in the odd-parity block. Therefore, on the $k_z = 0$ plane $p_x$ and $p_y$ orbits do not interact with $p_z$ orbit, a feature forbidding gap opening at band crossing. Next we show that the cubic symmetry of fcc lattice ensures band crossings. As shown in Fig. 2(a), $p_x$ orbit at *X* point $\boldsymbol{k} = \pi/a\,\hat{\boldsymbol{x}}$ and $p_z$ orbit at *Z* point $\boldsymbol{k} = \pi/a\,\hat{\boldsymbol{z}}$ are mapped into each other by a $\pi/2$ rotation around the *y* axis. The *Z* point at $\boldsymbol{k} = \pi/a\,\hat{\boldsymbol{z}}$ is connected to that at $\boldsymbol{k} = \pi/a\,\hat{\boldsymbol{x}} + \pi/a\,\hat{\boldsymbol{y}}$ by a reciprocal lattice vector of fcc lattice (see Fig. 2(a)), meaning that they are equivalent to each other. As the consequence, on the $k_z = 0$ plane $p_x$ orbit at *X* point should take the same energy as $p_z$ orbit at *Z* point. This makes a band crossing between $p_x$ and $p_z$ orbits unavoidable on any path confined in the $k_z = 0$ plane and connecting *X* and *Z* points. The same discussion applies for other high symmetry points and planes with $p_y$ orbit involved. Therefore, the symmetry of fcc lattice ensures the existence of NL.

It is heuristic to consider a tight-binding (TB) model with orbits hopping among nearest-neighboring sites of fcc lattice. In this case, we can derive analytically the trajectory of NL (see supplementary materials). For example, in the $k_z = 0$ plane, NL should obey

$$(1 - 4\tau^2) \cos(k_x a) \cos(k_y a) + \cos(k_x a) + \cos(k_y a) + 1 = 0 \quad (1)$$

with $\tau \equiv (t_0 + t_1)/2t_2$ where $t_0$ stands for the hopping integral between the same orbits residing on sites of different cubic planes, $t_1$ for that on the same cubic plane, and $t_2$ for that between different orbits (see supplementary materials). It is obvious that, independently of hopping integrals, Eq. (1) is satisfied at *W* points $\boldsymbol{k} = \pm\pi/a\,\hat{\boldsymbol{x}} \pm \pi/2a\,\hat{\boldsymbol{y}}$ and $\boldsymbol{k} = \pm\pi/2a\,\hat{\boldsymbol{x}} \pm \pi/a\,\hat{\boldsymbol{y}}$, and hence NL always pass through *W* points (see Fig. 2(a)). This is similar to the linear frequency dispersions at *K* points in 2D honeycomb photonic crystals [26].

The left-hand side of Eq. (1) changes sign for $\tau = 1$ at $\Gamma$ point, which divides the hopping parameter space into two regimes. For $|\tau| < 1$, the symmetry-guaranteed NL are shown in Fig. 2(a) in accordance to Eq. (1) and its counterparts for the other two mirror symmetric planes.

Dispersions at several typical cuts in the $k_z = 0$ plane are displayed in Fig. 2(b). Linear band crossings are clearly observed at *W* point and a point between *K* and Γ points inside the first BZ. For $|\tau| > 1$, in addition to the symmetry-guaranteed NL, there appear accidental NL around Γ point as shown in Figs. 2(c) and (d).

In terms of the reciprocal lattice vectors of fcc lattice, the NL including *W* points form a closed loop at *Z* point $\boldsymbol{k} = \pi/a\,\hat{\boldsymbol{z}}$ as depicted by the dashed lines in Figs. 2(a) and (c). In the whole reciprocal space, one finds a full network of symmetry-guaranteed NL as displayed in Fig. 3(a), similar to those revealed in a recent work [13], where NL are put as circles and accidental NL are not shown for clarity. The shape of NL changes with values of hopping integrals as displayed in Figs. 3(b-f), and both symmetry-guaranteed and accidental NL become straight and touch with each other for $|\tau| \gg 1$. As seen in Fig. 3(e), for $\tau = (\sqrt{2} - 1)/2$ the NL comes to *K* point, which is equivalent to *U* point $\boldsymbol{k} = \pi/a\,\hat{\boldsymbol{x}} + \pi/4a\,\hat{\boldsymbol{y}} + \pi/4a\,\hat{\boldsymbol{z}}$ (see Fig. 1(a)) [23]. The above discussions apply for the orbits $[d_{xy}, d_{xz}, d_{yz}]$ (see supplementary materials).

In reality electromagnetic fields are not confined well on the lattice sites as presumed in the TB model. In order to confirm the above results, we solve Maxwell's equations and check the harmonic modes of electromagnetic waves $\mathbf{E}(\boldsymbol{r}, t) = \mathbf{E}(\boldsymbol{r})e^{-i\omega t}$, $\mathbf{B}(\boldsymbol{r}, t) = \mathbf{B}(\boldsymbol{r})e^{-i\omega t}$ following the procedure provided in the section of methods. We choose silicon spheres with relative permittivity $\epsilon_{\text{fcc}} = 11.56$, which can be supported by dielectric foam with permittivity close to vacuum [22]. The frequency dispersions of an infinite system thus obtained are displayed in Fig. 4(a), where linear band crossings associated with topological NL are clearly observed at $f/f_0 \approx 0.29, 0.32$ and $0.35$ with $f_0 = c/2\pi a$.

Let us first focus on the high-frequency NL. Upon the spatial inversion the electric field is mapped by $\mathbf{E}(\boldsymbol{r}) \to -\mathbf{E}(-\boldsymbol{r})$, and thus one has $\mathbf{E}(\boldsymbol{r}) = \pm\mathbf{E}(-\boldsymbol{r})$ for odd- and even-parity states respectively. On the other hand, upon the mirror reflection with respect to the *yz* plane, the electric field is mapped by $\mathbf{E}(\boldsymbol{r}) \to M_{yz}\left(\mathbf{E}\left(M_{yz}(\boldsymbol{r})\right)\right) \equiv [-E_x(-x,y,z), E_y(-x,y,z), E_z(-x,y,z)]$, and the two other corresponding relations for the *xy* and *xz* planes, and thus one has $\mathbf{E}(\boldsymbol{r}) = \pm M_{\alpha\beta}\left(\mathbf{E}\left(M_{\alpha\beta}(\boldsymbol{r})\right)\right)$ for the even- and odd-parity states respectively. Checking the distribution of electric field upon these two symmetry operations, we find that the band carrying frequency $f/f_0 \approx 0.39$ at *X* point exhibits the symmetry same as $p_x$ orbit with the electric field at the center of sphere schematically shown in the left inset of Fig. 4(a). There are two degenerate bands at *X* point with frequency $f/f_0 \approx 0.32$, which are identified in the same way as $p_y$- and $p_z$-like orbits. Along the *XW* direction, the energy of $p_x$-like ($p_z$-like) band decreases (increases) and meet at *W* point to form the linear frequency dispersion at $f/f_0 \approx 0.35$. Therefore, these NL are well described by the TB model, even though electromagnetic waves distribute over the whole space, both inside and among the dielectric spheres.

The light cones over NL acquire a topological index since two bands of opposite parities with respect to the mirror reflection reverse their frequencies [10], in contrary to that at Γ point (see Figs. 2(a) and (c)). As an important consequence, novel 2D states should appear at the interface to a topologically trivial photonic insulator with photonic gap matching NL [1]. It is known that there is a global photonic gap in the diamond lattice of dielectric spheres [23]. For fcc lattice of silicon spheres, one can choose the diamond lattice of lead-glass spheres with the relative permittivity $\epsilon_{\text{dia}} = 5.78$ [27]. We consider the super lattice structure shown in Fig. 4(b) where a common lattice constant is taken for fcc and diamond lattices. In a slab infinite in the *xy* plane, only momenta $k_x$ and $k_y$ are good quantum numbers, and the 3D NL are projected to 2D ones as shown in inset of Fig. 4(c). The numerical results obtained by solving Maxwell's equation are displayed in Fig. 4(c), where a drumhead-shaped band is observed within the photonic gap which connects the two band crossing points denoted by the red dots. As shown in Fig. 4(d), the light associated with this new band is well localized the interface between fcc and diamond lattices. The light speed given by the group velocity of the in-gap band in Fig. 4(d) is small in this interface state, which can be used for achieving slow light yearned for many applications [17-19]. Within the projected NL around Γ point indicated in inset of Fig. 4(c), all states are gapped except for the interface one, indicating that it will not be influenced by other states.

In the same way, we find that NL around $f/f_0 \approx 0.29$ in Fig. 4(a) is achieved by *d*-like orbits, with distributions of electric magnetic field at the center of sphere schematically shown in the right inset of Fig. 4(a), which corresponds to the linear band crossings found in the previous work [23]. Interface states similar to that shown in Fig. 4(d) are also realized in this case. The *p*- and *d*-like bands appear in frequency bands well separated from each other near the surface of BZ, which makes it possible to treat them independently, jusfifying the TB model. There is an accidental NL around Γ point at $f/f_0 \approx 0.32$, where, in contrary to the symmetry-guaranteed NL, one cannot identify clearly *p*- and *d*-orbits.

NL can also be found in fcc-structured dielectric rods in air background [28], a fcc lattice of air spheres in the background of dielectric material [22], and fcc photonic structures written by femtosecond laser [29]. Our discussions apply also for phononic and electronic systems. Ignoring the spin-orbital coupling (SOC) in electronic systems, NL just become doublet owing to the spin degree of freedom. Taking into account SOC, the symmetry of fcc structure prohibits gap opening at *W* points, whereas gap will open at other points in BZ. In this case, the system becomes a Dirac semimetal [6], which still shows unique topological surface states.

**Methods**
**Solving Maxwell's equations:** For the present dielectric system, Maxwell's equations are simplified for harmonic modes into a master equation [16]

$$\frac{1}{\epsilon(\boldsymbol{r})} \boldsymbol{\nabla} \times \boldsymbol{\nabla} \times \mathbf{E}(\boldsymbol{r}) = \frac{\omega^2}{c^2} \mathbf{E}(\boldsymbol{r}) \qquad (2)$$

for the electric field, and the magnetic field is given by the Faraday relation $\mathbf{H} = -\frac{i}{\mu_0 \omega} \boldsymbol{\nabla} \times \mathbf{E}(\boldsymbol{r})$, where $\epsilon(\boldsymbol{r})$ is the position-dependent relative permittivity and $c = 1/\sqrt{\epsilon_0 \mu_0}$ is the light velocity in vacuum. Equation (2) can be solved numerically in momentum space using the package MIT PHOTONIC BANDS [30].

**Acknowledgements**

This work was supported by the WPI Initiative on Materials Nanoarchitectonics, Ministry of Education, Culture, Sports, Science and Technology of Japan, and Grants-in-Aid for Scientific Research (No.16K17755), JSPS.

**Figure Captions:**

**Figure 1 | FCC lattice of dielectric spheres and the first BZ. a,** Dielectric spheres are residing on the sites of the face-centered-cubic (fcc) lattice in the background of another dielectric material. **b**, The first Brillouin zone (BZ) with high symmetry points specified.

**Figure 2 | Distribution of NL in 3D momentum space and dispersion relations with linear band crossings. a** and **b,** NL (green curves) derived based on the TB model for the three $p$ orbits with nearest-neighbor hopping and dispersion relations along several momentum directions. All NL pass through $W$ points, and in terms of reciprocal lattice vectors NL form a loop shown by the dashed green curve around $Z$ point $\boldsymbol{k} = \pi/a\,\hat{\boldsymbol{z}}$. Linear band crossings are clearly found in the dispersion relations, where the red curve is for the band with large weight of $p_z$ orbit. Hopping parameters are $t_1 = 0.8t_0$ and $t_2 = 1.2t_0$ satisfying $\tau < 1$, and thus all the NL are guaranteed by the symmetries of fcc lattice. **c** and **d**, same as **a** and **b** except for $t_2 = 0.5t_0$ satisfying $\tau > 1$ where accidental NL appear additionally around $\Gamma$ point.

**Figure 3 | NL network in momentum space and dependence of NL shape on hopping integrals. a**, Network in the whole 3D momentum space formed by the symmetry-guaranteed NL. The loops of NL touch with each other at $W$ points denoted by white balls. For electronic systems with finite spin-orbital coupling, the energy degeneracy is only preserved by symmetry on $W$ points, yielding a Dirac-point semimetal. For clarity, the NL loops are put as circles. **b-f**, Possible configurations of NL shown in the $k_z = 0$ plane as governed by values of hopping integrals in terms of $\tau \equiv (t_0 + t_1)/2t_2$ derived in terms of Eq. (1) based on the TB model. Accidental NL shrinks to $\Gamma$ point for $\tau = 1$, and the symmetry-guaranteed NL come to $K$ and $U$ points for $\tau = (\sqrt{2} - 1)/2$.

**Figure 4 | Frequency dispersion relations for fcc lattice of silicon spheres and the interface states. a**, Frequency dispersion relations of fcc lattice of silicon spheres shown in Fig. 1(a) derived by solving the master equation (2). There are two sets of symmetry-guaranteed NL (denoted by open circles) at $f/f_0 \approx 0.29$ and $0.35$ associated with $d$- and $p$-orbits respectively, with distributions of electric field at the center of spheres displayed schematically in the insets (for details see the text), and one set of accidental NL (denoted by dots) at frequency $f/f_0 \approx 0.32$ where $f_0 = c/2\pi a$. The radius of spheres is $R = a/2$ and the relative permittivity is $\epsilon_{\text{fcc}} = 11.56$, whereas that of the background is taken as unity. **b**, Heterostructure of fcc and diamond lattices with the same lattice constant. For spheres in the diamond lattice lead glass with $\epsilon_{\text{dia}} = 5.78$ is chosen, which generates a photonic band covering the frequency band around that of the NL at $f/f_0 \approx 0.35$. The thickness of the fcc (diamond) lattice is of $34a$ ($16a$). **c,** Frequency dispersion relations along the direction of $k_y = 0$ for the structure shown in **b**, with the projected BZ and NL in the $k_x - k_y$ plane displayed in the inset. A drumhead-shaped band is observed within the photonic gap which connects the two band crossing points shown by the red dots. **d,** Intensity of the electric field in arbitrary unit on the cross section of the heterostructure indicated by dashed line in **a** for the drumhead-shaped band at $\Gamma$ point. It is found that light is well localized at the interface.

# Figures

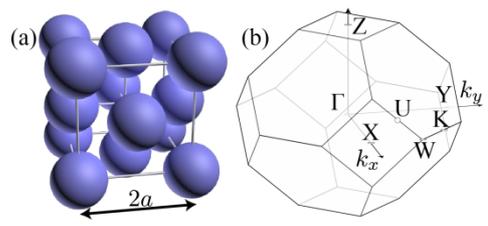

**Figure 1**

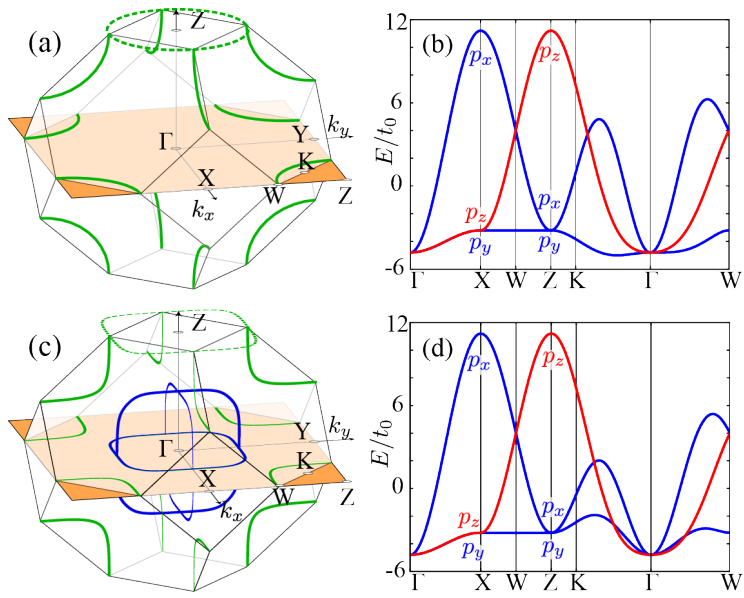

**Figure 2**

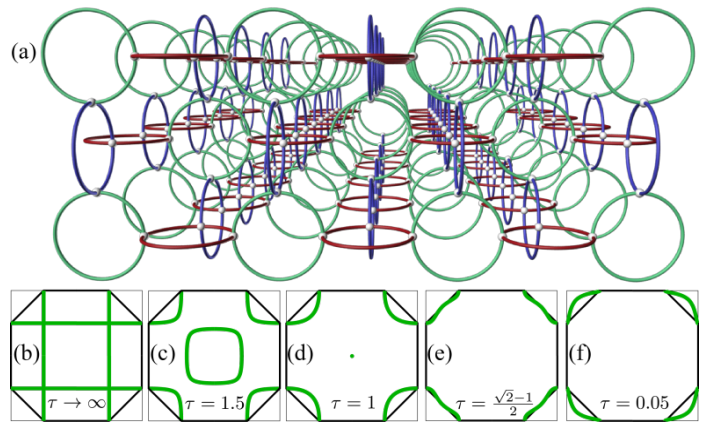

Figure 3

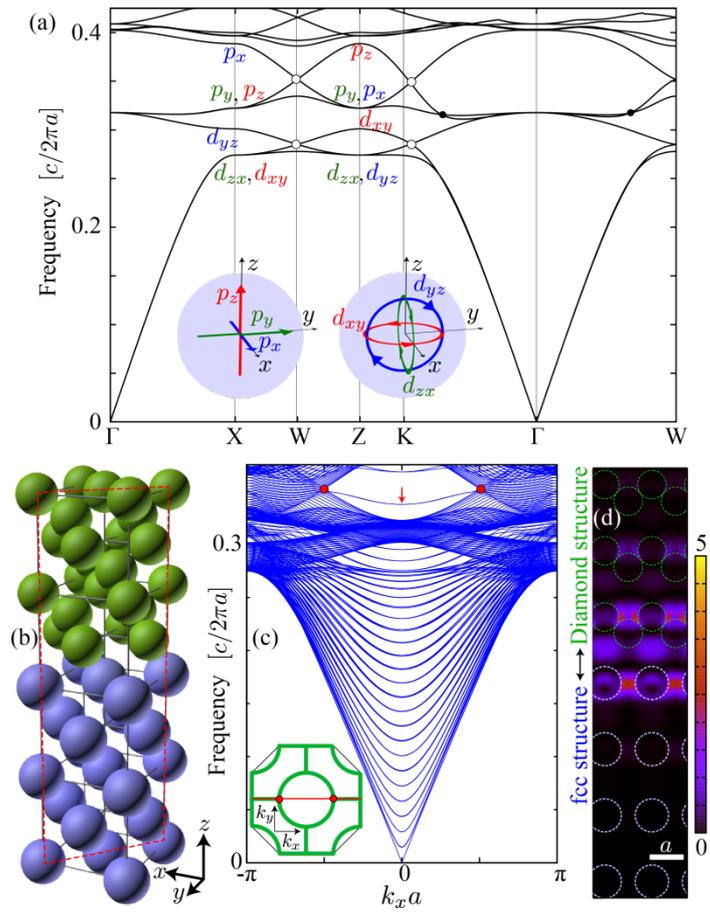

**Figure 4**